\documentclass[pra,twocolumn,showpacs]{revtex4-1}

\usepackage{graphicx,epstopdf}
\usepackage{amsmath}
\usepackage{amssymb}
\usepackage{color}
\usepackage{float}
\usepackage{epsfig}
\usepackage{epstopdf}
\usepackage{subfigure}
\usepackage{tcolorbox}

\usepackage[colorlinks=true, citecolor=blue, urlcolor=blue ]{hyperref}

\begin{document}

\title{ Tensor-network approach to compute genuine multisite entanglement\\ in  infinite quantum spin chains}

\author{Sudipto Singha Roy,\(^{1,2,3}\) Himadri Shekhar Dhar,\(^{4,5}\) Aditi Sen(De),\(^{3}\) and Ujjwal Sen\(^{3}\)}

\affiliation{\(^1\)Instituto de F{\'i}sica T{\'e}orica  UAM/CSIC, C/ Nicol{\'a}s Cabrera 13-15, Cantoblanco, 28049 Madrid, Spain
\\
\(^2\)Department of Applied Mathematics, Hanyang University (ERICA),
55 Hanyangdaehak-ro, Ansan, Gyeonggi-do, 426-791, Korea\\
\(^3\)Harish-Chandra Research Institute, HBNI, Chhatnag Road, Jhunsi, Allahabad 211 019, India \\
\(^4\)Institute for Theoretical Physics, Vienna University of Technology, Wiedner Hauptstra{\ss}e 8-10/136, A-1040 Vienna, Austria\\
\(^5\)Physics Department, Blackett Laboratory, Imperial College London, SW7 2AZ London, UK}

\date{\today}

\begin{abstract}

We devise a method based on the tensor-network formalism to calculate  genuine multisite entanglement in ground states of infinite spin chains containing  spin-1/2 or spin-1 quantum particles. The ground state is obtained by employing an infinite time-evolving block decimation method acting upon an initial matrix product state for the infinite spin system. We explicitly show how such infinite matrix product states with translational invariance provide a natural framework to derive the generalized geometric measure, a computable measure of genuine multisite entanglement, in the thermodynamic limit of quantum many-body systems with both spin-1/2 and  higher-spin particles.

\end{abstract}

\maketitle

\section{\label{sec1}Introduction}
In recent years,  entanglement~\cite{horodecki} has turned out to be an important characteristic in the study of low-dimensional strongly-correlated quantum systems, especially from the perspective of critical phenomena in the low-temperature regime of many quantum many-body systems~\cite{fazio-nature, nielsen,many-body,sir_maam} and implementation of quantum information protocols using solid-state, cold gas, and other  physical substrates~\cite{bose,briegel,kitaev,kane,whaley}. While most of the  attention in studying these systems has been bestowed on  bipartite entanglement measures such as entanglement of formation, concurrence, or block entanglement entropy, an important albeit difficult to estimate quantity is the multipartite entanglement in quantum many-body systems (see Ref.~\cite{horodecki}). Interestingly, it has often been observed that there exist some co-operative phenomena where bipartite entanglement and other known order parameters fail to detect the interesting physics, which are then captured by multipartite entanglement~\cite{manab_threebody, lin,anindya,orus,wei}. Moreover, the study of multiparty entanglement in  quantum systems with higher spins, even for finite-sized systems, remains largely unexplored.


When expanding the study of multipartite entanglement to understand  complex quantum phenomena in the thermodynamic limit, for both spin-1/2 and higher-spin quantum particles,  
the innate difficulty is to characterize computable entanglement measures
(for recent developments, see Refs.~\cite{manab_threebody,wei,orus,miranda,anindya,lin,our,shimony, goldbart, illuminati, sen}). In most instances,  for quantum many-body systems, the complexity in measuring multipartite entanglement scales exponentially  with  increasing dimension of the total Hilbert space, which in turn is associated with  the number of quantum systems involved in the problem, and can often be unamenable even with  approximate methods. In recent years, numerical techniques such as density matrix renormalization group (DMRG)~\cite{dmrg}, matrix product states (MPS)~\cite{mps}, and projected entangled pair states (PEPS)~\cite{rev-mps} have allowed unprecedented access to physical properties of many-body systems, including estimation of global entanglement in low-dimensional spin systems~\cite{wei,orus,lin}.
The growth of newer tensor-network methods~\cite{tn}, such as multi-scale entanglement renormalization ansatz (MERA)~\cite{mera}, along with other significant developments in higher-dimensional~\cite{peps} and topological quantum systems~\cite{topology}, provide newer directions to explore the role of multipartite entanglement in generic quantum systems.

In this work, we employ a tensor-network based approach to estimate the  genuine multipartite entanglement, which for pure quantum states characterizes the situation where  the many-body system cannot be formed by states that are product across some bipartition(s) of the multiparty  system.
We investigate this behavior in the thermodynamic limit of infinite chains of both spin-1/2 as well as spin-1 quantum systems. 
We show that matrix product states for infinite one-dimensional quantum spin systems, provide a natural framework to estimate the \textit{generalized geometric measure} (GGM)~\cite{sen} (see also~\cite{shimony, goldbart, illuminati}), which is a computable measure of genuine multipartite entanglement, defined by using the geometry of the space of multiparty states. 
To demonstrate the efficacy of our formalism, we first consider a set of prototypical Hamiltonians of low-dimensional quantum spin systems. For instance, we 
obtain the ground states for spin-1/2 systems such as the transverse Ising and the \emph{XYZ}  models, using infinite time-evolving block decimation ($i$TEBD)~\cite{vidal-itebd} of an initial state. We show how the GGM in  the thermodynamic limit of the system  can be estimated from the final infinite matrix product state ($i$MPS). Subsequently, we extend our study to  more complex  models  such as the spin-1 Ising model with transverse single-ion anisotropy. Here we observe that the genuine multipartite entanglement in the thermodynamic limit can clearly highlight the different quantum phases of the many-body system and the scaling of entanglement can identify the critical points.

The paper is arranged as follows. After the brief introduction in Sec.~\ref{sec1}, we discuss GGM as a measure of genuine multiparty entanglement in Section~\ref{sec2}. We then look at how expressions for the reduced states can be obtained from the infinite MPS picture in Sec.~\ref{sec3}. In Section~\ref{sec4} we look at how the ground states of spin chain models, containing spin-1/2 or spin-1 particles, can be derived using $i$TEBD. In Section~\ref{sec5}, we calculate the GGM for the ground states of these different models. We conclude with a discussion in Sec.~\ref{sec6}.

\section{\label{sec2}Generalized geometric measure}
A hierarchy of geometric measures of multiparty entanglement~\cite{illuminati} of an $N$-party pure quantum state, $|\Psi\rangle_N$, can be defined in terms of  geometric distance between the given state and the set of $k$-separable states, $\mathcal{S}_k$, which is the set of all pure quantum states that are separable across at least $k-1$ partitions in the system or alternatively, are a product of states of \(k\) subsystems.
%
Considering fidelity subtracted from unity, which is closely connected to the Fubini-Study and the Bures metrics, as our choice of distance measure, one can define the geometric measure of multiparty entanglement as~\cite{goldbart,shimony,illuminati,sen} 
\begin{equation}
{G_k}(|\Psi\rangle_N)=1-\max_{|\chi\rangle \in \mathcal{S}_k}|\langle \chi|\Psi\rangle_N|^2,
\label{GE}
\end{equation}
where $2 \leq k \leq N$ and $|\langle \chi|\Psi\rangle_N|^2$ is the fidelity. The maximization ensures that \(G_k\) measures how entangled (or far
away) a state \(|\Psi\rangle_N\) is  with respect to the (from the)
closest \(k\)-separable states. In principle, a set of $N-1$ measures of multipartite entanglement ($\{G_k\}$) can be defined, by employing the  minimum distances from the $N-1$ sets,  $\mathcal{S}_k$. Multipartite entanglement measures, such as the global entanglement~\cite{goldbart}, consider the distance of $|\Psi\rangle_N$ from the set of completely separable or $N$-separable states, $\mathcal{S}_{k=N}$. These measures do not detect separability that may occur across lesser number of partitions ($k < N$). A more stringent measure is the genuine multipartite entanglement, $G_{k=2}$, which corresponds to the minimum distance from the set $\mathcal{S}_{2}$. Since, $\mathcal{S}_k \subset \mathcal{S}_{k^\prime}$, if $k^\prime \leq k$, we get  $\mathcal{S}_{k} \subset \mathcal{S}_{2}~\forall~k$.
This implies that the minimal distance is computed by considering the minimization over all
\(k\)-separable states for all \(k\), and thus captures presence of genuine multiparty entanglement in the quantum state. In other words, non-zero value of $G_{k=2}$, in Eq.~(\ref{GE}), implies that $|\Psi\rangle_N$  is not separable across {\it any} bipartition. 

In general,  computation of $G_k$ appears to be hard, as it involves maximization over a large set of $k$-separable states.  Incidentally, $G_{k=2}$ for a quantum state, is equal to   the generalized geometric measure ($\mathcal{G}$)~\cite{sen}, which reduces  to
\begin{equation}
\mathcal{G}(|\Psi\rangle_N)=1-\max_{\mathcal{S}_{\mathcal{A}:\mathcal{B}}} \{ \lambda^{2}_{\mathcal{A}:\mathcal{B}} | \mathcal{A} \cup \mathcal{B} =\{N\}, \mathcal{A} \cap \mathcal{B} = \emptyset \},
\label{eqn:ggm}
\end{equation}
where $\mathcal{S}_{\mathcal{A}:\mathcal{B}}$ = $\mathcal{S}_{2}$, is the set of all bi-separable states, with bipartitions $\mathcal{A}$ and $\mathcal{B}$, and 
$\lambda_{\mathcal{A}:\mathcal{B}}$ = $\max\{\lambda_{\mathcal{A}:\mathcal{B}}^i\}$ for the Schmidt decomposition, $|\Psi\rangle_N=\sum_i \lambda^{i}_{\mathcal{A}:\mathcal{B}} |\phi^i\rangle_\mathcal{A} |\tilde{\phi}^i\rangle_\mathcal{B}$. Here $N$ denotes the set of $N$ parties possessing the state $|\Psi\rangle_N$.  The maximization over all $k$-separable states is reduced to optimization over the set of $\lambda^{2}_{\mathcal{A}:\mathcal{B}}$ across all possible bipartitions of $|\Psi\rangle_N$.   Such simplification of $\mathcal{G}(|\psi\rangle_N)$ helps to evaluate genuine multiparty entanglement content of multiparty  state involving arbitrary number of parties and in arbitrary dimensions. Note however that with increasing $N$, the number of possible choices of the  bipartitions also  increases exponentially, and hence  computation of $\mathcal{G}$  becomes cumbersome. In addition to this,  if the quantum state of the system  cannot be defined uniquely, it is also not possible to compute the value of $\mathcal{G}$ for the system. 

We present a brief outline of the proof for GGM ($\mathcal{G}$) being a measure of genuine multipartite entanglement \cite{sen} in the Appendix \ref{Appendix}. 
Importantly, we now show that the measure of $\mathcal{G}$  can be characterized in the language of tensor-network methods. In particular, we consider the MPS formalism in translationally invariant (TI) quantum systems, which provides a {natural framework} to estimate genuine multisite entanglement.

\section{\label{sec3}Analytical form  of reduced density matrices}
The maximum Schmidt coefficient across a bipartition required for  GGM is the square-root of the maximum eigenvalue of the reduced density matrix of the subsystems across the bipartition. Obtaining the reduced density matrices of an infinite-sized system, using the MPS formalism, is the  primary motivation of the paper. Let us begin with the preliminary MPS representation of a  many-body quantum state, $|\Psi\rangle_N$,  given by \cite{rev-mps,tn}
\begin{eqnarray}
\label{psi_mps}
|\Psi\rangle_N &=& \sum_{i_1i_2\dotsi_N}\sum_{\alpha_2\dots \alpha_{N-1}} \text{Tr} (A^{i_1}_{\alpha_1,\alpha_2}A^{i_2}_{\alpha_2 \alpha_3}\dots A^{N}_{\alpha_{N-1}\alpha_N})\nonumber\\
&\times& |i_1, i_2, i_3, \dots i_N\rangle,
\end{eqnarray}
where $i_k$ is  the physical index, with the local system dimension $d$, and $\alpha_k$ being the auxiliary index, each with a bond dimension $D$. 
$\{A^{i_k}\}$ are thus $D \times D$ matrices corresponding to each $k$ site.
For low values of $D$, the MPS representation of $|\Psi\rangle_N$ is very efficient as the number of parameters required to express the state scales with $N$ as $ND^2d$, instead of  $d^N$. This can be further reduced by considering some potential symmetry in the system, such as translational invariance of $\{A^{i_k}\}$ matrices. Importantly, in order to obtain the reduced density matrices of a quantum many-body  system, one should be able to efficiently compute the $\{A^{i_k}\}$ matrices. However, there are only a few cases for which the exact MPS form of the quantum state is known \cite{rev-mps,tn}. One such example is the unnormalized  $N$-qubit Greenberger-Horne-Zeilinger (GHZ)~\cite{ghz} state,  
$|\textrm{GHZ}\rangle_N$ = $|0\rangle^{\otimes N}+|1\rangle^{\otimes N}$, which  is  local unitarily equivalent to the possible entangled ground state of the Ising chain at large coupling strength~\cite{ghz_gs}. For $D$ = 2 (and $d$ = 2 for qubits), the matrices for the MPS are  $\{A^{i_k}\} = \{A^0(k), A^1(k)\}$ = $\{\sigma^+ \sigma_x, \sigma^- \sigma_x\}$, $\forall~k$, where $\sigma_k$s are the usual Pauli matrices, and $\sigma^{\pm}=\frac{1}{2}(\sigma_x\pm i\sigma_y)$. We note that the $A^{i_k}$ matrices are translationally invariant. Another example of TI systems is the ground state of the AKLT Hamiltonian~\cite{AKLT}, where for $d$ = 3 and $D$ = 2, $\{A^{i_k}\} = \{A^0(k), A^1(k), A^2(k)\}$ = $\{\sigma_z, \sqrt{2}\sigma^+, -\sqrt{2} \sigma^-\}$, $\forall~k$. 
However, in general,  
the matrices $\{A^{i_k}\}$ can
have explicit site dependence. For example, consider the $N$-qubit $W$-state~\cite{W_state},  $|\textrm{W}\rangle_N$=$\frac{1}{\sqrt{N}}(|10\dots0\rangle+|01\dots0\rangle+\dots|00\dots1\rangle)$,  which is known to be the ground state of the ferromagnetic \textit{XX} model with strong transverse field. Interestingly, although the state is translationally invariant,  the $\{A^{i_k}\}$ matrices are not, as shown  for $D$ = 2. Here, $\{A^0(k), A^1(k)\}$ = $\{{\sigma^+}, \mathbb{I}_2\}$, for $k < N$, and 
$\{A^0(k), A^1(k)\}$ = $\{\sigma^+ \sigma_x, \sigma_x\}$, for $k = N$, where  $\mathbb{I}_2$ is the $2\times 2$ identity matrix \cite{garcia}.

In general, for MPS with site-dependent $\{A^{i_k}\}$, calculation of reduced density matrices of quantum states beyond moderate-sized systems may require considerable computational effort, especially if the bond dimension $D$ is not small. This is a significant road-block in the computation of GGM. However, if the system is TI, i.e. $\{A^{i_k}\} = A^i, \forall ~k$, and the  $A^i$ matrices  can be efficiently estimated, then the reduced density matrices can be obtained even for infinite sized systems, thus allowing us to compute the genuine multipartite entanglement of quantum states in the thermodynamic limit.
Let us begin with an MPS representation of a TI quantum system with local dimension $d$ and $\{A^i\}$, with bond dimension $D$. The MPS could be obtained as a ground state of a physical Hamiltonian or a time-evolved quantum state, quenched from some initial product state. 
To calculate the reduced density matrices, we first consider the case for single-site reduced state first 
from the multi-qubit TI  MPS.
For a very small system-size, viz. $N=2$, and known  $A^i$ matrices, the expression for the single-site reduced density matrix, is given by
$
\rho^1=\frac{1}{E^2}
\text{tr}((A^0\otimes \bar{A}^0)E)|0\rangle\langle 0|+\text{tr}((A^0\otimes \bar{A}^1)E)|0\rangle\langle 1|\nonumber + \text{tr}((A^1\otimes \bar{A}^0)E)|1\rangle\langle 0|+\text{tr}((A^1\otimes \bar{A}^1)E)|1\rangle\langle 1|,
$
where $E=\sum_i A^i\otimes \bar{A}^i$ is the transfer matrix of the translationally invariant system and $\bar{A}$ is the conjugate transpose of $A$. Similarly, for $N=3$,
$
\rho^1=\frac{1}{E^3}
\text{tr}((A^0\otimes \bar{A}^0)E^2)|0\rangle\langle 0|+\text{tr}((A^0\otimes \bar{A}^1)E^2)|0\rangle\langle 1|\nonumber + \text{tr}((A^1\otimes \bar{A}^0)E^2)|1\rangle\langle 0|+\text{tr}((A^1\otimes \bar{A}^1)E^2)|1\rangle\langle 1|.\nonumber
$
For an arbitrary $N$ and local dimension, $d$ = 2 (qubit),  the expression for the single-site density matrix is given by
\begin{eqnarray}
\rho^1&=& \sum_{i,j = 0}^{1}  \frac{\text{tr}((A^i \otimes \bar{A}^j)E^{N-1})}{E^N}~|i\rangle\langle j|.
\label{single-site}
\end{eqnarray}
At this stage, our aim is to generalize Eq.~(\ref{single-site}) for very large, and eventually, infinite systems. To this end,  we first consider the  spectral decomposition of the transfer matrix, in the MPS formalism for infinite system, known   as $i$MPS, $E^N=\sum_i\lambda_i^N |L^i\rangle \langle R^i|$, where $|L^i\rangle$ and $|R^i\rangle$ are the left and right eigenvectors, respectively. For  $N\rightarrow \infty$,  $E$ has 1 as a non-degenerate eigenvalue and all other eigenvalues have modulus smaller than 1, i.e.
$
E^N= |L^0\rangle\langle R^0|+\sum_{j=2}^{D^2}\lambda^N_k |L^k\rangle\langle R^k|.
$
Hence, as $N\rightarrow \infty$, $E^N \rightarrow  |L^0\rangle\langle R^0|$.
Thus, the elements of $\rho^1$, as expressed in Eq.~(\ref{single-site}), are given by
\begin{equation}
\rho^1_{ij}=\frac{\langle L^0|A^i\otimes \bar{A}^j|R^0\rangle}{\langle L^0|R^0\rangle}.
\end{equation}

Similarly, one can obtain the form of all $m$-consecutive site $l,l+1,l+2\dots$  ($m \geq2$)  reduced density matrices, using the relation
\begin{eqnarray}
\rho^m_{ij}=\frac{\langle L^0|A^{i_1} A^{i_2}\dots A^{i_m}\otimes \bar{A}^{j_1}\bar{A}^{j_2}\dots \bar{A}^{j_m}|R^0\rangle}{ \langle L^0|R^0\rangle},
\label{two_site}
\end{eqnarray}
where $i=i_1 i_2\dots i_m$ and $j=j_1 j_2\dots j_m$. 
For non-consecutive sites, $l,l+r_1,l+r_1+r_2,\dots$ the expression of the $m$-site reduced density matrix is given by 
\begin{eqnarray}
\rho_{ij}^m=\frac{\langle L^0|\tilde{A}^1E^{r_1-1} \tilde{A}^2 E^{r_2-1}..\tilde{A}^m|R^0\rangle }{ \langle L^0|R^0\rangle},\nonumber
\end{eqnarray} 
where $\tilde{A}^k=(A^{i_k}\otimes \bar{A}^{j_k})$.

This has remarkable significance as the number of parameters required to represent the $m$-site density matrices is reduced from $d^{m}$ to $D^2d$.
The reduced density matrix can thus be used to estimate the genuine multisite entanglement in systems described using infinite MPS.

\section{Ground state MPS using $i$TEBD \label{sec4}}

We briefly describe the algorithm to simulate the ground state of an infinite, one-dimensional quantum many-body Hamiltonian, $\mathcal{H}$, using the infinite MPS formalism. We start with an arbitrary MPS, $|\Psi\rangle_N$, as expressed in Eq.~(\ref{psi_mps}), and then eventually build the ground state $i$MPS using infinite time-evolving block decimation method.
To this end, starting from $|\Psi\rangle_N$, we perform an imaginary time-evolution:  $ |\Psi\rangle_N\rightarrow e^{-\tau \mathcal{H}}|\Psi\rangle_N$. The ground state configuration $|\Psi_0\rangle_N$ is then  obtained when $\tau$ becomes very large i.e. $|\Psi\rangle_N\sim|\Psi_0\rangle_N+\sum_{i=1}^{d^N} e^{-\tau (E_i-E_0)}|\Psi_i\rangle_N\xrightarrow{\tau\rightarrow\infty}|\Psi_0\rangle_N.$   In order to perform the $i$TEBD, we first use  second order Suzuki-Trotter (ST) decomposition~\cite{suzuki_trotter} on the exponential unitary operation and express each term in the TI matrix product operator (MPO) form~\cite{pirvu,pirvu_thesis}. 
This  essentially helps to change the optimization  problem of the  energy for the total system,  to the optimization associated with each decomposed TI MPO.  After one such ST iteration, we obtain an  MPS, $|\Psi^t\rangle$, which, in general, has a bigger bond dimension than the initial MPS. Therefore, one needs to truncate this to the allowed bond dimension $D$. We then normalize the imaginary time evolved state and choose that as a seed for the next time iteration. After each such ST step,  energy per site ($E_0/N=\frac{1}{N} \langle \Psi^t|\mathcal{H}|\Psi^t\rangle$) is calculated and  the expressions of the $\{A^i\}$ matrices for the $i$MPS of the ground state of the given Hamiltonian are then obtained by minimizing the energy. In general, energy per site scales with the size of the system. However, through some intermediary steps,  one can show that for $N\rightarrow \infty$,  it converges, to $E_{\infty}$ (say). Hence,  the final $\{A^i\}$ matrices are obtained when the energy per site converges.

To apply the above $i$MPS formalism we begin with a one-dimensional quantum system consisting of spin-1/2 particles. Such a quantum many-body Hamiltonian  can be written, with a certain degree of genericity, as
\begin{eqnarray}
\label{eqn:parent}
\mathcal{H}=\sum_{\langle ij \rangle}\big(J_x S^x_iS^x_{j}+J_y S^y_iS^y_{j}+\Delta  S^z_iS^z_{j}\big)+\sum_ihS^z_i,~~
\end{eqnarray}
where $J_x$, $J_y$ are the coupling constants along $x$- and $y$- directions respectively, $\Delta$ is the ``anisotropy"  along the $z$- direction, $h$ is the strength of the transverse field, $S^k=\sigma_k$ are the Pauli spin matrices, and $\langle ij \rangle$ denotes  the nearest-neighbor sites. Two important models that can be derived from $\mathcal{H}$ are the transverse {Ising} (in the limit  $J_y$ = $\Delta$ = 0), {the anisotropic \textit{XYZ} model ($J_{x(y)} = J
\pm\gamma$, and $h=0$)~\cite{sachdev1,sondhi,phase_2}. Note that in the limit 
$\gamma=0$, the \emph{XYZ} model reduces to the  anisotropic \emph{XXZ} model, which  has gained some attention in studies on strongly-correlated systems~\cite{xxz}.
We note that in recent years, cooperative phenomena in quantum spin chains have been widely explored in the context of quantum information theory, especially in terms of entanglement ~\cite{fazio-nature, nielsen,many-body,sir_maam} and other quantum correlations~\cite{vedral_rmp_qc}.

We next look at the $i$MPS representation for  more complex quantum spin systems. For instance, we consider a quantum many-body chain with higher-spin particles, viz. the spin-1 Ising model with a transverse field akin to parameters arising from single-ion anisotropy generated by crystal fields~\cite{spin1:refs}. These systems can also be considered to be a derivative of the Blume-Emery-Griffiths model~\cite{begmodel}, where the quadratic terms have been neglected.
Such models have lately been used to study phase transitions in multicomponent fluids and semiconductor systems~\cite{beg2}.
The Hamiltonian of the spin-1 model is thus given by
\begin{equation}
\bar{\mathcal{H}}=\mathcal{J}_z \sum_{\langle i j\rangle} \mathcal{S}_i^z  \mathcal{S}_j^z+ \mathcal{K} ( \mathcal{S}_i^x)^2,
\label{Ham_spin1}
\end{equation}
where  $\mathcal{S}^i$'s are generalizations of the Pauli matrices for a spin-1 system, $\mathcal{J}_z$ denotes the coupling along the $z$-direction and $\mathcal{K}$ denotes the strength of the single-ion anisotropy parameter due to the crystal field in the transverse direction. The model undergoes a quantum phase transition at $\frac{\mathcal{J}_z}{\mathcal{K}}=2$~\cite{spin1:refs}.

In implementing the $i$MPS form and the $i$TEBD algorithm for obtaining the ground state of these Hamiltonians, we fix the bond dimension at $D=10$ and choose the initial Trotter step  to be $\tau=10^{-2}$, which is then gradually changed to $10^{-6}$ to improve accuracy. The convergence of the ground state energy is determined with an accuracy  $10^{-6}$. Once the ground state $i$MPS is obtained, one can access the Schmidt coefficients across all possible bipartitions of the quantum state by contracting the tensors efficiently, as shown in Eq.~(\ref{two_site}). The behavior of genuine multisite entanglement in ground state phases of the  Hamiltonian,  in the thermodynamic limit, can then be estimated from the generalized geometric measure.

\begin{figure}[t]
\epsfig{figure = 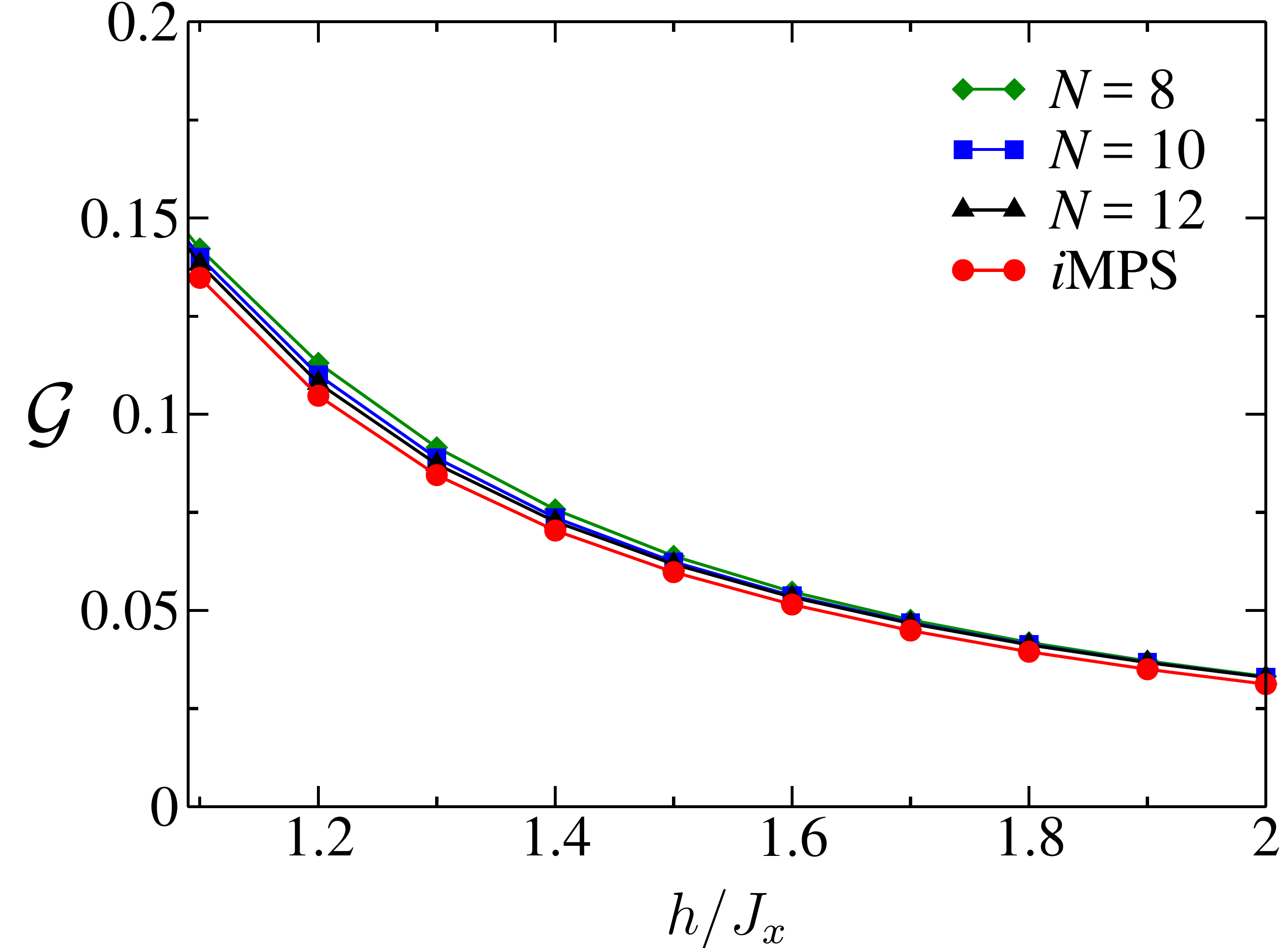, width=0.4\textwidth,angle=-0}
\caption{(Color online.) Variation of GGM ($\mathcal{G}$) with field strength ($h$) for the transverse Ising model, for different one-dimensional lattice sizes, viz. $N$ = 8 (green-diamonds), 10 (blue-squares), 12 (black-triangles) and infinite $N$ (red-circles). Both axes represent dimensionless quantities.}
\label{fig:ggm_ising}
\end{figure}

\section{\label{sec5}Genuine multisite entanglement in the thermodynamic limit}

For transverse Ising model, we consider a region away from critical point ($h/J_x = 1$), viz.~$1.1\leq h/J_x\leq2$.  The variation of GGM ($\mathcal{G}$) with respect to the transverse field strength $h/J_x$, is depicted in Fig.~\ref{fig:ggm_ising}. 
The thermodynamic limit of the genuine multisite entanglement, in the infinite spin lattice, is compared with the corresponding values obtained for finite-sized lattices ($N$ = 8, 10, and 12) using exact diagonalization. In order to compute the value of GGM ($\mathcal{G}$) using  exact diagonalization method, in all the cases ($N=8, 10, 12$), we perform the optimization in Eq.~(\ref{eqn:ggm}),  by taking into account   all possible  bipartitions (whose number  is $\sum_{i=1}^{6} {{N}\choose{i}}$ for $N=12$). We note that for the transverse field Ising model, maximum value of Schmidt coefficient always comes from the single-site reduced density matrices. We use this fact to compute the value of GGM ($\mathcal{G}$) in the thermodynamic limit using $i$MPS. Therefore,  in our case, Eq.~(\ref{single-site}) will serve the purpose. For this model, in the region parametrized by $0\leq {h}/{J_x}\leq 0.8$,  energy gap closes and as discussed earlier, it is not possible to compute the multiparty entanglement using the measure GGM for non-unique ground states. The figure shows a distinct scaling of $\mathcal{G}$ at field strengths closer to the critical point, $h/J_x = 1$. In this region,  difference between the GGM ($\mathcal{G}$) values, obtained  using exact diagonalization method ($N=12$) and $i$MPS, turns out to be at most  $\approx 10^{-3}$.  Away from it, $\mathcal{G}$ quickly becomes scale invariant, and approaches its thermodynamic limit even for low $N$. Here, difference between the GGM ($\mathcal{G}$) values computed  for $N=12$ and $i$MPS, becomes $\lesssim 10^{-4}$. 

\begin{figure}[t]
\epsfig{figure = 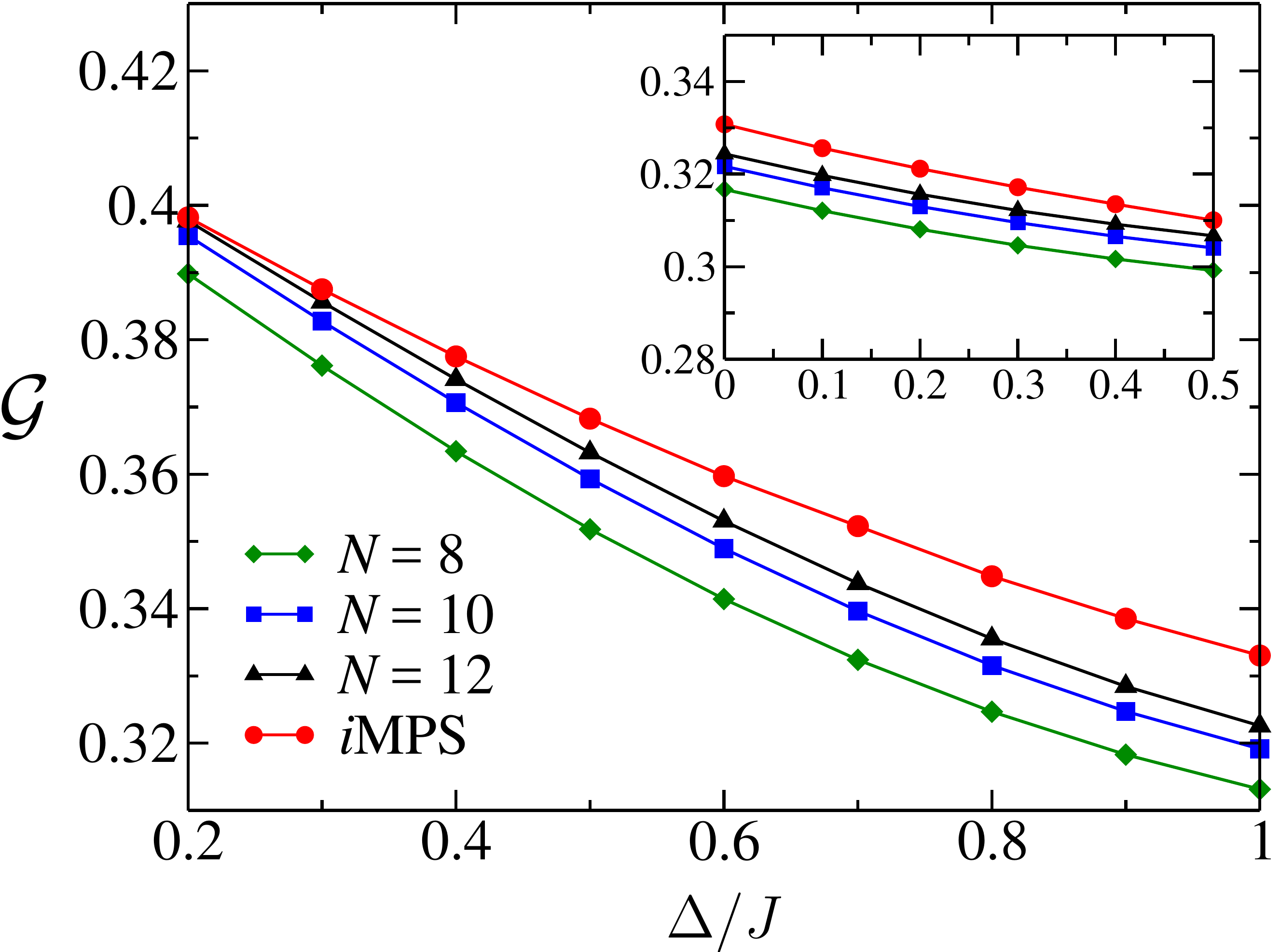, width=0.4\textwidth,angle=-0}
\caption{(Color online.) Variation of GGM ($\mathcal{G}$) with $\Delta/J$ for the \emph{XYZ} model  with $\gamma=0.5$, for different one-dimensional lattice sizes. Both axes represent dimensionless quantities. In the inset we plot the same quantities for $\gamma=0$ case (\emph{XXZ} model).}
\label{fig:ggm_xyz}
\end{figure}

Let us now consider the \emph{XYZ} Hamiltonian in absence of magnetic field, i.e., $J_x, J_y,\Delta \neq0$, $h=0$.
The behavior of $\mathcal{G}$ with $\Delta/J$, for the anisotropic \emph{XYZ} Hamiltonian with $\gamma=0.5$, is depicted in Fig.~\ref{fig:ggm_xyz}.  Unlike the Ising case, from the exact diagonalization results for this model, we note that the maximum value of Schmidt coefficient always comes from the consecutive two-site reduced density matrices. We  again use this result to compute the value of GGM in the thermodynamic limit using $i$MPS. Therefore,  in this case, we use  Eq.~(\ref{two_site}) for computation of the maximum Schmidt coefficients. Like as the Ising case. here also degeneracy hinders us to find a unique ground state for the region $-1\leq \Delta/J\leq0$. Therefore, for this model, we consider following region between two critical points, for both finite and infinite lattices, parametrized by  $0.2 \leq \Delta/J \leq1.0$.  
Figure~\ref{fig:ggm_xyz}  shows that in contrast to the transverse Ising model, no scale invariance is achieved for $\mathcal{G}$ even away from the critical points, and it is not possible to achieve the thermodynamic limit by exactly diagonalizing  a spin model with  small system size.  In this case, difference between the GGM values computed  for $N=12$ and $i$MPS, at small values of $\Delta/J$ becomes $\lesssim 10^{-3}$, which further increases to $\lesssim 10^{-2}$ as $\Delta/J$ tends to 1.
For the \emph{XXZ} model ($\gamma=0$) (see the inset of Fig.~\ref{fig:ggm_xyz}), where the critical points are known to exist in the vicinity of $\Delta/J =\pm 1$, a similar absence of scale invariance is observed.  Here   difference between the GGM values computed  for $N=12$ and $i$MPS never decreases below $10^{-3}$.
Thus, $i$MPS plays a significant role in computing genuine multipartite entanglement in these systems.

\begin{figure}[h]
\epsfig{figure = 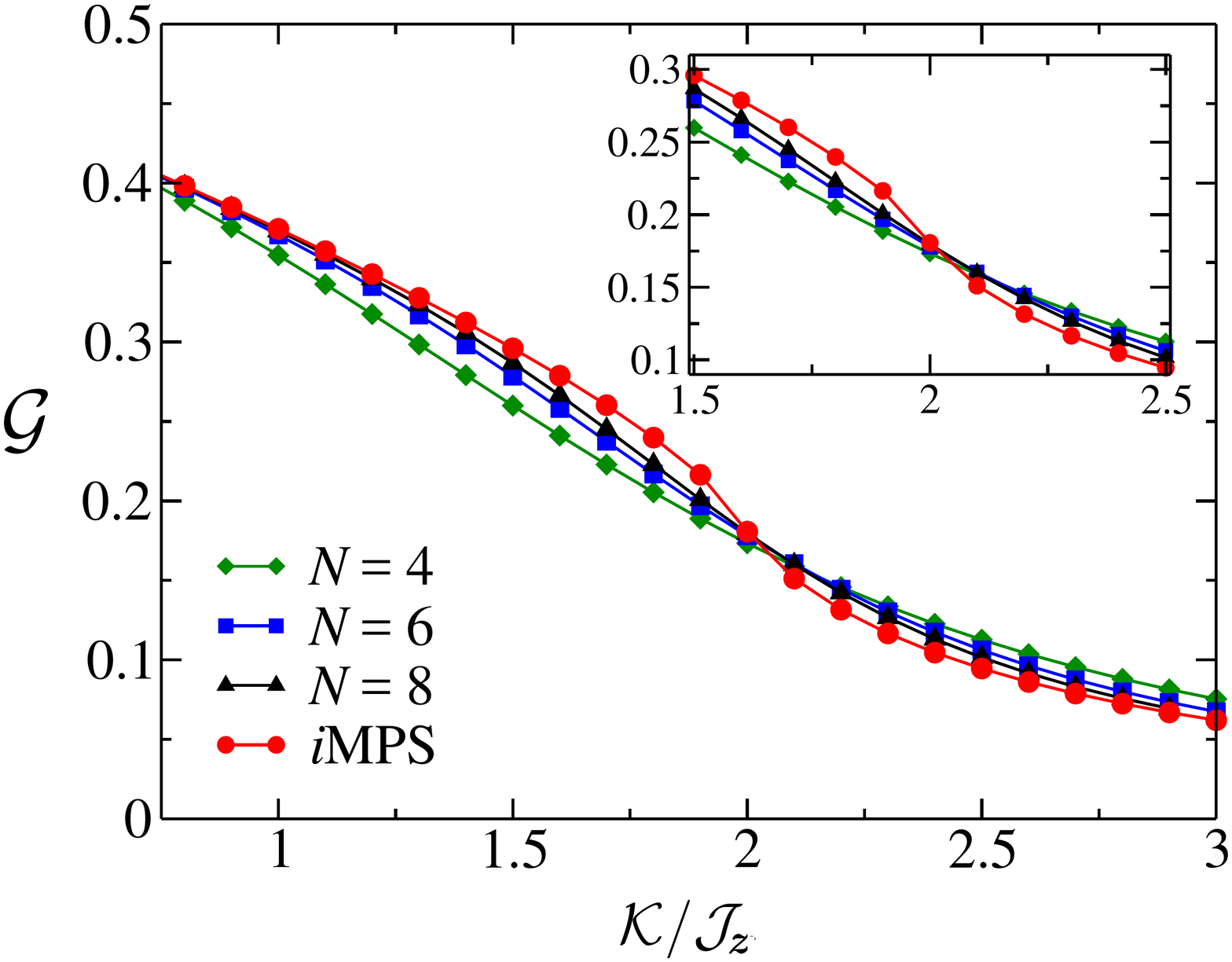, width=0.45 \textwidth,angle=-0}
\caption{(Color online.) Variation of GGM ($\mathcal{G}$) with  $\frac{\mathcal{K}}{\mathcal{J}_z}$ for the spin-1 model described in Eq.~(\ref{Ham_spin1}), for different one-dimensional lattice sizes, viz. $N$ = 4 (green-diamonds), 6 (blue-squares), 8 (black-triangles), and infinite $N$ (red-circles). Both axes represent dimensionless quantities. In the inset, we show the GGM in the region close to the transition point  ($\frac{\mathcal{K}}{\mathcal{J}_z} \approx 2$).}
\label{fig:ggm_ising_spin1}
\end{figure}

We now look at the genuine multipartite entanglement properties of the more complex higher-spin model, viz. the spin-1 Ising model with single-ion anisotropy as expressed in Eq.~(\ref{Ham_spin1}). We note that in contrast to the spin-1/2 models,  behavior of multipartite entanglement in this spin-1 model is unexplored even for finite spin systems. Here, we look at the behavior of GGM in the thermodynamic limit of the system using the $i$MPS formalism. 
The behavior of genuine multiparty entanglement is plotted in Fig.~\ref{fig:ggm_ising_spin1}. We again note that like  the transverse Ising model, the maximum Schmidt coefficient in this case also comes from the single-site reduced density matrices. Moreover, as in the previous cases, we observe that GGM starts decreasing monotonously with the increase of the strength of the single-ion anisotropy or the crystal field. However, for the spin-1 model, the scaling pattern of GGM in the thermodynamic limit shows several interesting features. For instance, before $\mathcal{K}<2$, in most of the regions, GGM increases with system size. On the other hand, for $\mathcal{K}>2$, the trend is reversed, i.e., the value of GGM decreases with the  increase of $N$. 
However, the variation of GGM with the anisotropy parameter clearly detects the critical points in the system. 
We observe that near the value $\mathcal{K}\approx 2$, GGM becomes almost scale invariant, which is 
a known value at which quantum phase transition occurs in the system. Therefore, our study shows that the scaling of GGM can identify the vital characteristics of the critical phenomena in the spin-1 model.

\section{\label{sec6}Discussion} In this work, we have shown how the tensor-network approach provides a natural structure to study  genuine multiparty entanglement, quantified by generalized geometric measure, in many-body quantum systems. In particular, the method involved  matrix product states to efficiently obtain the reduced density matrices of infinite quantum spin lattices, which upon making use of symmetries such as translational invariance of the matrices, allowed us to accurately estimate the generalized geometric measure of systems consisting of both spin-1/2 and higher-spins. The method thus provided us a viable  theoretical framework to look at interesting cooperative and critical phenomena by investigating multiparticle physical quantities in the thermodynamic limit of quantum many-body systems.

Importantly, this approach to compute generalized geometric measure using tensor networks is in principle also applicable for higher-dimensional lattices, provided the relevant tensors under the $i$MPS formalism are accessible using available numerical techniques.
Finally, we also note that the formalism presented in the work may provide useful directions in investigating genuine multipartite entanglement properties in several quantum systems, including condensed matter, photonic, and other topological systems, where tensor-network methods have turned out  to be  successful in studying  physical properties.

\acknowledgments
The research of SSR was supported in part by the INFOSYS scholarship for senior students.
HSD acknowledges funding by the Austrian Science Fund (FWF), project no.~M 2022-N27, under the Lise Meitner programme of the FWF.

\appendix
\section{Proof of GGM as a measure of genuine multiparty entanglement }
\label{Appendix}
Here, we present a very concise proof for the GGM to be a measure of
genuine multiparty entanglement, starting from the concept of $k$-separability and the definition of the geometric measures of multiparty entanglement in Eq.~(\ref{GE}).
An important point to note is that $G_2$ is the minimum distance from the set of all $k$-separable quantum states, $\mathcal{S}_k \forall ~k$. {However, in principle,} as measurements  over general entangled bases yield higher or equal values as compared to those over product bases,  the maximum fidelity in Eq.~(\ref{GE}), can always be considered  from the set $\mathcal{S}_k$ with lowest $k$, as they contain more clustered partitions. Hence, for 
 $G_2$, the  set $\mathcal{S}_2$  of bi-separable states contains a closest separable state. 
Let  $\{\lambda_{\mathcal{A}:\mathcal{B}}^i\}_{i = 1}^d$ and $\{|\phi^i\rangle_\mathcal{A}, |\tilde{\phi}^i\rangle_\mathcal{B}\}_{i = 1}^d$ be the set of real, non-negative Schmidt coefficients and corresponding orthogonal vectors, respectively, across the bipartition $\mathcal{A}:\mathcal{B}$, where $d$ = $\max\{d_\mathcal{A},d_\mathcal{B}\}$. A bi-separable state, in general, can be written as, $|\chi\rangle$ = $|\eta\rangle_{\mathcal{A}}|\tilde{\eta}\rangle_{\mathcal{B}}$. 
The fidelity is then given by 
\begin{eqnarray}
|\langle \chi|\Psi\rangle_N| &=& |\sum_i \lambda^{i}_{\mathcal{A}:\mathcal{B}} \langle\eta|\phi^i\rangle_\mathcal{A} \langle\tilde{\eta}|\phi^i\rangle_\mathcal{B}| \nonumber\\
&=& |\sum_i \lambda^{i}_{\mathcal{A}:\mathcal{B}} f^i_\mathcal{A} ~g^i_\mathcal{B}|.
\end{eqnarray} 
A value of fidelity, possibly non-maximal, corresponds to  $|\eta\rangle_{\mathcal{A}}$ = $|\phi^k\rangle_\mathcal{A}$ and $|\tilde{\eta}\rangle_{\mathcal{B}} $ =  $|\tilde{\phi}^k\rangle_\mathcal{B}$, such that $f^k_\mathcal{A}  = g^k_\mathcal{B}$ = 1, where $k$ gives $\lambda_{\mathcal{A}:\mathcal{B}}$ = $\lambda^k_{\mathcal{A}:\mathcal{B}}$ =  $\max\{\lambda_{\mathcal{A}:\mathcal{B}}^i\}$. Thus we have, $|\langle \chi|\Psi\rangle_N| \geq \lambda_{\mathcal{A}:\mathcal{B}}$. 
However,  
\begin{equation*}
|\langle \chi|\Psi\rangle_N| \leq \sum_i \lambda^{i}_{\mathcal{A}:\mathcal{B}} |f^i_\mathcal{A}| |g^i_\mathcal{B}| \leq \lambda_{\mathcal{A}:\mathcal{B}}\sum_i |f^i_\mathcal{A}| |g^i_\mathcal{B}| \leq 
\lambda_{\mathcal{A}:\mathcal{B}}, 
\end{equation*}
where we have used the triangle-law for absolute values, and the relations, $ \sum_i \lambda_{\mathcal{A}:\mathcal{B}} \geq  \sum_i \lambda^i_{\mathcal{A}:\mathcal{B}}$ and $\sum_i |f^i_\mathcal{A}| |g^i_\mathcal{B}| \leq 1$. This gives us the desired relation, $|\langle \chi|\Psi\rangle_N|$ = $\lambda_{\mathcal{A}:\mathcal{B}}$ = $\max\{\lambda_{\mathcal{A}:\mathcal{B}}^i\}$. For all bi-separable states, $|\chi\rangle$, $\lambda_{\mathcal{A}:\mathcal{B}}$ = 1, and as expected $\mathcal{G}(|\Psi\rangle_N) = 0$.
From, Eq.~(\ref{eqn:ggm}), we see that the maximum among the real and positive Schmidt coefficient squared, across all possible bipartitions, subtracted from unity gives the GGM, which measures the genuine multipartite entanglement in the system.

\end{document}